\documentclass{article}

\usepackage{epsfig} 
\newlength{\abstractwidth} 
\renewcommand{\thefootnote}{\fnsymbol{footnote}} 
\renewcommand{\thanks}[1]{\footnote{#1}} 
\newcommand{\starttext}{ 
\setcounter{footnote}{0} 
\renewcommand{\thefootnote}{\arabic{footnote}}} 
\renewcommand{\theequation}{\thesection.\arabic{equation}} 
\newcommand{\bea}{\begin{eqnarray}} 
\newcommand{\eea}{\end{eqnarray}} 
\newcommand{\beq}[1]{\begin{equation} \label{#1}} 
\newcommand{\be}{\begin{equation}} 
\newcommand{\ee}{\end{equation}}

\def\12{{1 \over 2}} 
\newcommand{\half}{{1\over 2}}

\begin{document} 
\renewcommand{\theequation}{\thesection.\arabic{equation}} 
\bigskip
\centerline{\Large \bf {A Model of Unified Gauge Interactions}}
\bigskip
\begin{center} 
{\large James Lindesay\footnote{ 
Address: Department of Physics, Howard University, Washington, DC 20059, 
email: jlindesay@howard.edu}
} \\
Computational Physics Laboratory \\
Howard University
\end{center}
\bigskip\bigskip 
\begin{abstract} 

Linear spinor fields are a generalization of the Dirac field that have direct correspondence
with the known physics of fermions, inherent causality properties in their most fundamental constructions,
and positive mass eigenvalues for all particle types.  
The algebra of the generators for infinitesimal transformations of these fields
directly constructs the Minkowski metric \emph{within} the internal group space as a 
consequence of non-vanishing commutation relations between generators that carry space-time indexes.
In addition, the generators have a fundamental matrix representation
that includes Lorentz transformations within a group that unifies internal gauge symmetries generated by a set of hermitian
generators for SU(3)$\times$SU(2)$\times$U(1), and nothing else.  The construction of linearly independent
internal SU(3) and SU(2) symmetry groups necessarily involves the
mixing of three generations of the mass eigenstates labeling the (massive) representations of the linear spinor fields.  The group
algebra also provides a mechanism for the dynamic mixing of massless particles of differing  ``transverse mass" eigenvalues
conjugate to the affine parameter labeling translations along their
light-like trajectories.  The inclusion of a transverse mass generator is necessary for group closure of the extended Poincare algebra,
but its eigenvalue must vanish for massive particle representations. 
A unified set of space-time group transformation operations
along with internal gauge group symmetry operations for linear spinor fields will be demonstrated
in this paper.

\end{abstract} 

\starttext \baselineskip=17.63pt \setcounter{footnote}{0} 

\setcounter{equation}{0}
\section{Introduction}
\indent

The Dirac equation utilizes a matrix algebra to construct a 
field equation that is linear in the quantum operators for 4-momentum. 
Since inversion of linear operations is straightforward, 
the properties of evolution dynamics described using such linear operations on quantum states have direct interpretations
(e.g. towards constructions of resolvants or propagators)\cite{JLFQG}\cite{LMNP}\cite{AKLN}.
The Dirac formulation can be extended to
generally require that the form $\hat{\Gamma}^\mu \: \hat{P}_\mu$
be a Lorentz scalar operation, resulting in
a spinor field equation of the form
\be
\mathbf{\Gamma}^\beta \cdot {\hbar \over i} { \partial \over \partial x^\beta} \,
\hat{\mathbf{\Psi}}_{(\gamma)}^{(\Gamma)}
(\vec{x}) = -(\gamma) m c  \, \hat{\mathbf{\Psi}}_{(\gamma)}^{(\Gamma)}(\vec{x}) ,
\label{LinearConfigurationSpinorFieldEqn}
\ee
where $m$ is positive for all particle types, and $\mathbf{\Gamma}^\beta$ are
finite dimensional matrix representations of the operators $\hat{\Gamma}^\beta$. 
For massive particles, the particle type label $(\gamma)$ is just the particular eigenvalue of the hermitian matrix $\mathbf{\Gamma}^0$.
In the $\Gamma=\half$ representation, the matrix form of the
operators $\mathbf{\Gamma}^\beta= {1 \over 2} \mathbf{\gamma}^\beta$ are one half of the Dirac matrices\cite{Dirac}\cite{BjDrell},
and the particle type label takes values $(\gamma) = \pm \half$, eliminating the need for any 
filled ``Dirac sea" of negative energy states.

\section{An Extension of the Lorentz and Poincare Groups} 
\indent 

An extension of the Lorentz group can be defined using the algebra
\bea
\left [ \Gamma^0 \, , \, \Gamma^k \right] \: = \: i \, K_k  ,\\
\left [ \Gamma^0 \, , \, J_k \right] \: = \: 0  ,\\
\left [ \Gamma^0 \, , \, K_k \right] \: = \: -i \,  \Gamma^k  ,\\
\left [ \Gamma^j \, , \, \Gamma^k \right] \: = \: -i \, \epsilon_{j k m} \, J_m  ,\\
\left [ \Gamma^j \, , \, J_k \right] \: = \: i \, \epsilon_{j k m} \, \Gamma^m  ,\\
\left [ \Gamma^j \, , \, K_k \right] \: = \: -i \, \delta_{j k} \, \Gamma^0  ,
\label{ExtLorentzGroupEqns}
\eea
which has a Casimir operator $
C_\Gamma \: = \: \underline{J} \cdot \underline{J} \,-\, \underline{K} \cdot \underline{K}
\,+\, \Gamma^0 \, \Gamma^0 \,-\, \underline{\Gamma} \cdot \underline{\Gamma}$
that commutes with all generators of the group. 
The operators $C_\Gamma$,  $\Gamma^0$, and  $J_z$ have been chosen as the set of mutually
commuting operators for the construction of the finite dimensional representations.

The matrices corresponding to $\Gamma={1 \over 2}$ 
(the fundamental representation)
 have dimensionality $N_{1 \over 2}=4$, and
will be expressed in terms of the Pauli spin matrices $\sigma_j$ as demonstrated below:
\be
\begin{array}{ll}
\mathbf{\Gamma^0} \,=\, {1 \over 2} \left( \begin{array}{cc}
\mathbf{1} & \mathbf{0} \\ \mathbf{0} & -\mathbf{1} \end{array} \right)
 \quad \quad &
\mathbf{J}_j \,=\, {1 \over 2} \left( \begin{array}{cc}
\sigma_j & \mathbf{0} \\ \mathbf{0} & \sigma_j 
\end{array} \right) \\ \\
\mathbf{\Gamma}^j \,=\, {1 \over 2} \left( \begin{array}{cc}
\mathbf{0} & \sigma_j \\ 
-\sigma_j & \mathbf{0} \end{array} \right) &
\mathbf{K}_j \,=\, -{i \over 2} \left( \begin{array}{cc}
\mathbf{0} & \sigma_j \\ 
\sigma_j & \mathbf{0} \end{array} \right) 
\end{array}
\label{4x4RepresentationEqn}
\ee
A representation for the $\Gamma=1$ matrices can be found in Appendix D.2.1 of
reference \cite{JLFQG}.

The inclusion of space-time translations into the group algebra
must result in a self-consistent closed set of generators.
The 4-momentum operators
together with the extended Lorentz group operators
do not produce a closed group structure, due to Jacobi relations of the
type $[\hat{P}_j , [\hat{\Gamma} ^ 0 , \hat{\Gamma} ^k] ]$.  
An additional momentum-like operator, which will be labeled $\mathcal{M}_T$,
must be introduced, resulting in additional non-vanishing commutators:
\bea
\left [ J_j \, , \, P_k \right] \: = \: i \hbar \, \epsilon_{j k m} \, P_m ,
\label{JPeqn} \\
\left [ K_j \, , \, P_0 \right] \: = \: -i \hbar \, P_j ,
\label{KP0eqn} \\
\left [ K_j \, , \, P_k \right] \: = \: -i \hbar \, \delta_{j k} \, P_0 ,
\label{KPeqn} \\
\left [ \Gamma^\mu \, , \, P_\nu \right] \: = \: i \, \delta_\nu ^\mu \, \mathcal{M}_T  c,
\label{GamPeqn} \\
\left [ \Gamma^\mu \, , \, \mathcal{M}_T \right] \: = \: {i \over c} \, \eta^{\mu \nu} \, P_\nu ,
\label{GamGeqn}
\eea
The final two relations extend the Poincare algebra as necessary to close the algebra.

Linearity of the operator $\Gamma^\mu P_\mu$ 
in the energy-momentum generators is quite useful in developing the field equations
defining linear spinor fields. 
Important commutation relations of this operator are given below:
\be
\left [ J_k, \Gamma^\mu P_\mu \right ] \: = \: 0
\label{JDirac}
\ee
\be
\left [ K_k, \Gamma^\mu P_\mu \right ] \: = \: 0
\label{KDirac}
\ee
\be
\left [ P_\beta, \Gamma^\mu P_\mu \right ] \: = \: 
-i \mathcal{M}_T P_\beta
\label{PDirac}
\ee
\be
\left [ \mathcal{M}_T, \Gamma^\mu P_\mu \right ] \: = \: 
-i \eta^{\beta \nu} P_\beta P_\nu
\label{MTDirac}
\ee
From (\ref{MTDirac}), the operator $\hat{\mathcal{M}}_T$
only commutes with
$\Gamma^\mu P_\mu$ for massless particles,
while from (\ref{PDirac}) the 4-momentum operator
only commutes with
$\Gamma^\mu P_\mu$ for those states with eigenvalues of $\hat{\mathcal{M}}_T$ that vanish. 
For massless states $m_T \neq 0$,
the operator $\hat{\mathcal{M}}_T$ is the generator for translations along the affine parameter labeling the
light-like trajectory of the massless particle.

\subsection{Unitary massive particle states}

A Casimir operator for the complete extended Poincare (EP) group 
whose eigenvalues label irreducible particle states can be constructed
from the Lorentz invariants
\be
\mathcal{C}_m \: \equiv \: \mathcal{M}_T^2 c^2 \,-\, \eta^{\beta \nu} P_\beta P_\nu .
\label{Casimir_mu}
\ee
This form suggests that the
hermitian operator $\mathcal{M}_T$ be referred to as a
\emph{transverse mass} parameter of the state.
The quantum (standard) state vectors labeled using mutually commuting operators
satisfy
\be
\begin{array}{l}
\hat{\mathcal{C}}_m \, \left | m, \Gamma,  \gamma, J, s_z \right > ~=~
m^2 c^2 \, \left | m, \Gamma,  \gamma, J, s_z \right > , \\
\hat{C}_\Gamma \,\left | m, \Gamma,  \gamma, J, s_z \right > ~=~
2 \Gamma (\Gamma + 2) \, \left | m, \Gamma,  \gamma, J, s_z \right > , \\
\hat{\Gamma}^0 \, \left | m, \Gamma,  \gamma, J, s_z \right > ~=~
\gamma \, \left | m, \Gamma,  \gamma, J, s_z \right >, \\
\hat{J}^2 \, \left | m, \Gamma,  \gamma, J, s_z \right > ~=~
J(J+1) \hbar^2  \,\left | m, \Gamma,  \gamma, J, s_z \right >, \\
\hat{J}_z \, \left | m, \Gamma,  \gamma, J, s_z \right > ~=~
s_z \hbar \, \left | m, \Gamma,  \gamma, J, s_z \right >,
\end{array}
\label{StateVectorEqn}
\ee
where $\Gamma$ is an integral or
half-integral label of the representation of the extended Lorentz
group, and $J$, which has the same integral signature as $\Gamma$,
 labels the internal angular momentum representation of the state.
Unitary representations of general momentum states are generated via
boosting standard states satisfying (\ref{StateVectorEqn}).  
For a more complete treatment of the algebra, symmetries, and causality properties
of linear spinor fields, the reader is invited to examine sections 4.3 and 4.4 in
reference \cite{JLFQG}.

\subsection{Development of group metric on space-time indexes}

For a general algebra satisfying
$\left [ \hat{G}_r \, , \, \hat{G}_s \right ] \: = \: -i \, \sum_m \left ( c_s \right ) _r ^m \, \hat{G}_m $,
the adjoint representation expressed in terms of the structure constants defines a group metric $\eta_{a b}$
given by 
\be
\eta_{a b} \: \equiv \: \sum_{s \, r} \left ( c_a \right )_r ^s \, \left ( c_b \right )_s ^r .
\ee
This group metric defines invariants on products of group generators, such as the 
Casimir operator.  In particular, for
the non-commuting operators $\Gamma^\mu$ that carry a space-time index
conjugate to the 4-momentum, a group metric describing an invariance is given by
\be
\eta^{(EP)} _{\Gamma^\mu \, \Gamma^\nu} \: = \: 8 \, \eta_{\mu \, \nu}
\ee
where $\eta_{\mu \, \nu}$ is the usual Minkowski metric of the Lorentz group. 
The Minkowski metric is  thus non-trivially generated explicitly within the extended Lorentz group algebra
(beyond the Lorentz invariance \emph{implicit} in Lorentz transformations). 
This \emph{group theoretic} metric can be used to develop Lorentz invariants using the
operators $\Gamma^\mu$, which transforms as a contravariant 4-vector operator. 
Its explicit use in group invariants directly connect group operations to curvilinear coordinate transformations. 
One should note that the group operator $\mathcal{I} \equiv \eta^{(EP)} _{\Gamma^\mu \, \Gamma^\nu} \Gamma^\mu \, \Gamma^\nu$ is not
proportional to the identity matrix, or even a diagonal matrix, for general representations $\Gamma>{1 \over 2}$. 
However, $\mathcal{I}$ commutes with all generators of Lorentz group transformations,
as does  $\Gamma^\mu P_\mu$.

The standard Poincare group has no non-commuting operators
that can be used to explicitly connect the group structure to the metric properties of space-time
translations.
Since the generators $P_\mu$ transform as covariant 4-vectors under arbitrary coordinate transformations
(of which group transformations are a specific subset), the group structure generating the linear spinor
fields is explicitly tied to curvilinear space-time dynamics through the principle of equivalence.

\section{A Complete Set of Hermitian Generators}
\indent

The fundamental representation of the extended Lorentz group can be developed in terms
of $4 \times 4$ matrices of the group GL(4). 
Since there are 16 Hermitian generators whose representations are  $4 \times 4$ matrices,  there are
an additional 12 Hermitian generators in the group\cite{JLLSF13}.  Of course, one of those generators is proportional to the identity
matrix,  and it generates
a U(1) internal abelian symmetry group defining a conserved hypercharge on the algebra.

One can construct a linearly independent set of 11 additional generators using
the Hermitian forms of the anti-Hermitian generators
$\mathbf{\Gamma}^j$ and $\mathbf{K}_j$ given by
$\mathbf{T}_j = i \,  \mathbf{\Gamma}^j$ and 
$\mathbf{T}_{j+3} = i \, \mathbf{K}_j$,
 two additional independent generators given by
\be
\mathbf{T}_{7} ={i \over 2} \left (
\begin{array}{cc}
\mathbf{0} & \mathbf{1} \\
-\mathbf{1} & \mathbf{0}
\end{array}
\right )
\quad , \quad
\mathbf{T}_{8} ={1 \over 2} \left (
\begin{array}{cc}
\mathbf{0} & \mathbf{1} \\
\mathbf{1} & \mathbf{0}
\end{array}
\right )  \, ,
\ee
and a final set of three generators $\mathbf{T}_9, \, \mathbf{T}_{10},$ 
and $\mathbf{T}_{11}$ forming a closed representation of SU(2) on the lower components:
\be
\mathbf{T}_{j+8} ={1 \over 2} \left (
\begin{array}{cc}
\mathbf{0} & \mathbf{0} \\
\mathbf{0} & \mathbf{\sigma}_j
\end{array}
\right ) ,
\ee
where  $\mathbf{0}$ are
$2 \times 2$ zero matrices.
Although they are linearly independent of all other generators in the unified group,
the set of 8 Hermitian generators $\mathbf{T}_s$ for $s:1 \rightarrow 8$
do not form a closed algebra independent of the other Hermitian
generators.  However, of course, the group of all 15 traceless generators close within
the algebra of SU(4).

 In the constructions that follow, the internal group symmetries will
initially be developed for a masive particle state transformed 4-spinor of the form
 $\bar{\psi} (x) = \left (  \begin{array}{c}\bar{ \phi}_1 (x) e^{i \bar{\omega}_1(x)}  \\ 0 \\ 0 \\ 0  \end{array} \right )$. 
While the generators $\mathbf{T}_9, \, \mathbf{T}_{10},$ and $\mathbf{T}_{11}$ leaves the transformed state
spinor $\bar{\psi}$  invariant, there is no combination of the generators $\mathbf{T}_s$  
for $s:1 \rightarrow 8$ that does so.
However, one can construct a set of linearly independent generators of an SU(3) algebra that \emph{can} be transformed
into an internal symmetry through CKM mixing between three generations. 
In what follows, a straightforward set of generators for internal SU(2) and SU(3) symmetries will be constructed
to demonstrate the unified closure of a group of 11 linearly independent hermitian generators, along with the 4 hermitian generators in the
extended Lorentz group defining the representations of the linear spinor fields.

\subsection{Invariance of an internal SU(2) algebra}

The generators $\tau_j \equiv \mathbf{T}_{j+8}$ transform under a set of internal tranformations $\mathbf{M}^{(2)}$ in the space of reduced dimension that leaves
the transformed spinor $\bar{\psi}$ unchanged, in the form
\be
\mathbf{M}^{(2)}= \left (
\begin{array}{cc}
\mathbf{1} & \mathbf{0} \\
\mathbf{0} & \mathbf{S}^{(2)}
\end{array}
\right ) ,
\label{Eq:M2}
\ee
where $\mathbf{S}^{(2)}$ is a unitary unimodular transformation matrix in SU(2), and $\mathbf{1}$ is the 2$\times$2 identity matrix. 
All generators transform in this reduced dimensional subspace according to
$\mathbf{G}'_r = \mathbf{M}^{(2)} \mathbf{G}_r (\mathbf{M}^{(2)})^{-1}$, preserving their group algebra.  The transformed state
spinor $\bar{\psi}$ is invariant under transformations involving $ \mathbf{M}^{(2)}$, making it an internal invariance group for this spinor,
since any SU(2) rotated set of the generators could alternatively have been chosen to construct this independent subspace
without altering any of the extended Lorentz group generators.

It is clear that the eigenbasis of this chosen set of independent SU(2) transformations is the same as that of the generators
$\mathbf{\Gamma}^0$ and $\mathbf{J}_3$ that describe (little group) invariance transformations on the particle
mass states.  This means that transitions that are purely induced by this SU(2) symmetry will preserve internal
quantum numbers within any single mass eigenstate generation of $\bar{\psi}$.  However, there is no
additional set of linearly independent SU(3) transformations in this basis, requiring that any mix of SU(2) and
SU(3) induced transformations necessarily involves mixing among eigenbases.  This mixing will be demonstrated
in the next section.

\subsection{Construction of an internal SU(3) algebra}

To begin, consider the following set of SU(3) generators  that leave $\bar{\psi}$ invariant:
\begin{displaymath}
\mathbf{t}_1 = \left (
\begin{array}{cccc}
0 & 0 & 0 & 0 \\
0 & 0 & 0 & 0 \\
0 & 0 & 0 & {1 \over 2} \\
0 & 0 & {1 \over 2} & 0 
\end{array}
\right ) , \quad
\mathbf{t}_2 = \left (
\begin{array}{cccc}
0 & 0 & 0 & 0 \\
0 & 0 & 0 & 0 \\
0 & 0 & 0 & {i \over 2} \\
0 & 0 & -{i \over 2} & 0 
\end{array}
\right ) , \quad
\mathbf{t}_3 = \left (
\begin{array}{cccc}
0 & 0 & 0 & 0 \\
0 & 0 & 0 & 0 \\
0 & 0 & -{1 \over 2} & 0 \\
0 & 0 & 0 & {1 \over 2} 
\end{array}
\right ) , 
\end{displaymath}
\begin{displaymath}
\mathbf{t}_4 = \left (
\begin{array}{cccc}
0 & 0 & 0 & 0 \\
0 & 0 & 0 & {1 \over 2} \\
0 & 0 & 0 & 0 \\
0 & {1 \over 2} & 0 & 0 
\end{array}
\right ) , \quad
\mathbf{t}_5 = \left (
\begin{array}{cccc}
0 & 0 & 0 & 0 \\
0 & 0 & 0 & {i \over 2} \\
0 & 0 & 0 & 0 \\
0 & -{i \over 2} & 0 & 0 
\end{array}
\right ) , \quad
\mathbf{t}_6 = \left (
\begin{array}{cccc}
0 & 0 & 0 & 0 \\
0 & 0 & {1 \over 2} & 0 \\
0 & {1 \over 2} & 0 & 0 \\
0 & 0 & 0 & 0 
\end{array}
\right ) , 
\end{displaymath}
\be
\mathbf{t}_7 = \left (
\begin{array}{cccc}
0 & 0 & 0 & 0 \\
0 & 0 & {i \over 2} & 0 \\
0 & -{i \over 2} & 0 & 0 \\
0 & 0 & 0 & 0 
\end{array}
\right ) , \quad
\mathbf{t}_8 = \left (
\begin{array}{cccc}
0 & 0 & 0 & 0 \\
0 & -1 & 0 & 0 \\
0 & 0 & {1 \over 2} & 0 \\
0 & 0 & 0 & {1 \over 2} 
\end{array}
\right ) .
\ee
The generators $\mathbf{t}_j$ define a set of internal tranformations $\mathbf{M}^{(3)}$ in the space of reduced dimension of the form
\be
\mathbf{M}^{(3)}(\underline{\alpha})= \left (
\begin{array}{cc}
1 & \mathbf{0}^T \\
\mathbf{0} & \mathbf{S}^{(3)}
\end{array}
\right ) =e^{i \sum \alpha^s \mathbf{t}_s},
\ee
where $\mathbf{S}^{(3)}$ is a unitary unimodular transformation matrix in SU(3) and $\mathbf{0}$ is a 1$\times$3 zero vector. 
The transformed state spinor $\bar{\psi}$ is invariant under transformations involving 
$ \mathbf{M}^{(3)}$, making $ \mathbf{M}^{(3)}$ an internal invariance group for the spinor.
The SU(3) `flavor' eigenstates will be defined using this basis.

It is important to note that the matrices $\mathbf{t}_s$ do not form a set of linearly independent generators in the 
particle state representation eigenbasis
for which $\mathbf{\Gamma}^0$ and $\mathbf{J}_3$ are diagonal (in part because there can only be
3 independent diagonal traceless generators total).  A construction of independent generators in
the particle state eigenbasis requires that three `generations' of `flavor' eigenstates be mixed from the reduced subspace. 
A convenient mechanism for developing the appropriate mixing
is provided through general CKM matrices\cite{Cabibbo,KMmix} embedded within GL(4).
The particular choice for mixing will be 
the set of all transformations on the $3 \times 3$ subspace that leaves the causal partner of
$\bar{\psi}$ invariant demonstrated  below:
\begin{displaymath}
\mathbf{M}_{23} = \left (
\begin{array}{cccc}
1 & 0 & 0 & 0 \\
0 &  \cos(\theta_{23}) & \sin(\theta_{23}) & 0 \\
0 &  -\sin(\theta_{23}) & \cos(\theta_{23}) & 0  \\
0 & 0 & 0 & 1 
\end{array}
\right ) , \quad
\mathbf{M}_{31} = \left (
\begin{array}{cccc}
\cos(\theta_{31}) & 0 &  \sin(\theta_{31})e^{- i \delta_{31}} & 0 \\
0 & 1 & 0 & 0 \\
- \sin(\theta_{31})e^{i \delta_{31}} & 0 &  \cos(\theta_{31}) & 0  \\
0 & 0 & 0 & 1 
\end{array}
\right ) , 
\end{displaymath}
\be
\mathbf{M}_{12} = \left (
\begin{array}{cccc}
\cos(\theta_{12})  & \sin(\theta_{12}) & 0 & 0 \\
 -\sin(\theta_{12}) &  \cos(\theta_{12})   & 0 &0 \\
0 & 0 & 1 & 0 \\
0 & 0 & 0 & 1
\end{array}
\right ) , \quad
\mathbf{U}_{CKM} \equiv \mathbf{M}_{23} \mathbf{M}_{31}\mathbf{M}_{12} .
\ee
Most CKM transformations of this type on the generators $\tilde{\mathbf{t}}_s \equiv \mathbf{U}_{CKM} \mathbf{t}_s \mathbf{U}_{CKM}^1$
will produce a set of generators $\tilde{\mathbf{t}}_s$ satisfying the algebra of SU(3) that, along with three additional generators, form
a group of 11 linearly independent hermitian generators alternative to the previous set $\left \{  \mathbf{T}_1, ... , \mathbf{T}_{11}  \right  \}$.

For clarity, consider the example CKM transformation from the SU(3) eigenbasis to the $\mathbf{\Gamma}^0, \mathbf{J}_3$ eigenbasis
parameterized using
($\theta_{12}\rightarrow 0, \theta_{23}\rightarrow 0, \theta_{31}\rightarrow {\pi \over 4}, \delta_{31} \rightarrow 0$):
\be
\mathbf{U}_{CKM} = \left (
\begin{array}{cccc}
{1 \over \sqrt{2}} & 0 & {1 \over \sqrt{2}} &  0 \\
0 & 1 & 0 & 0 \\
-{1 \over \sqrt{2}} & 0 & {1 \over \sqrt{2}} & 0 \\
0 & 0 & 0 &  1 
\end{array}
\right )  .
\ee
The prior set of 11 linearly independent hermitian generators
$\left \{  \mathbf{T}_1, ... , \mathbf{T}_{11}  \right  \}$
can be directly decomposed in terms of a new set of 11
linearly independent hermitian generators given by
\begin{displaymath}
\left \{  \tilde{\mathbf{t}}_1, ... , \tilde{\mathbf{t}}_8 ,
 \tilde{\mathbf{\Delta}}_1 \equiv (\mathbf{T}_1-\mathbf{T}_5)/2, 
\tilde{\mathbf{\Delta}}_2  \equiv (\mathbf{T}_2+\mathbf{T}_4)/2, 
\tilde{\mathbf{\Delta}}_3  \equiv (\mathbf{T}_3+\mathbf{T}_7)/2  \right \}.
\end{displaymath}
The set of matrices $\{ \tilde{\mathbf{t}}_1, ... , \tilde{\mathbf{t}}_8  \}$ continue to obey the
same closed SU(3) algebra as the transformed generators $\{ \mathbf{t}_1, ... , \mathbf{t}_8   \}$ in the internal SU(3) eigenbasis.
To demonstrate their linear independence in the unified algebra of the linear spinor fields, the original set of 11 hermitian matrices
are decomposed in Eqn. \ref{Eq:SU3decomp}.
\be
\begin{array}{lll}
\mathbf{T}_1 = -\mathbf{J}_2 - \sqrt{2}\, \tilde{\mathbf{t}}_2 + \sqrt{2}\, \tilde{\mathbf{t}}_7 + 2 \tilde{\mathbf{\Delta}}_1 & 
\mathbf{T}_2 = \mathbf{J}_1 - \sqrt{2}\, \tilde{\mathbf{t}}_1 - \sqrt{2}\, \tilde{\mathbf{t}}_6 + 2 \tilde{\mathbf{\Delta}}_2 &
\mathbf{T}_3 =  -\tilde{\mathbf{t}}_5 +  \tilde{\mathbf{\Delta}}_3 \\
\mathbf{T}_4 = -\mathbf{J}_1 + \sqrt{2}\, \tilde{\mathbf{t}}_1 + \sqrt{2}\, \tilde{\mathbf{t}}_6  & 
\mathbf{T}_5 = -\mathbf{J}_2 - \sqrt{2}\, \tilde{\mathbf{t}}_2 + \sqrt{2}\, \tilde{\mathbf{t}}_7  &
\mathbf{T}_6 =  -\mathbf{J}_3 - {3 \over 2}\tilde{\mathbf{t}}_3 - \tilde{\mathbf{t}}_4 +  {1 \over 2} \tilde{\mathbf{t}}_8 \\
\mathbf{T}_7 =  \tilde{\mathbf{t}}_5 +  \tilde{\mathbf{\Delta}}_3 &
\mathbf{T}_8 =  -\mathbf{J}_3 - {3 \over 2}\tilde{\mathbf{t}}_3 + \tilde{\mathbf{t}}_4 +  {1 \over 2} \tilde{\mathbf{t}}_8 \\
\mathbf{T}_9 = \sqrt{2}\, \tilde{\mathbf{t}}_1 - \tilde{\mathbf{\Delta}}_2 & 
\mathbf{T}_{10} =- \sqrt{2}\, \tilde{\mathbf{t}}_2 + \tilde{\mathbf{\Delta}}_1 &
\mathbf{T}_{11} ={1 \over 2} ( -\mathbf{\Gamma}^0 + \mathbf{J}_3 - \tilde{\mathbf{t}}_3 -\tilde{\mathbf{t}}_8 ) 
\end{array}
\label{Eq:SU3decomp}
\ee
The previously discussed 3 internal SU(2) generators $\tau_j \equiv \mathbf{T}_{j+8}$ are no longer in the set of 11 linearly independent hermitian generators.
One sees that any relationship between the `flavor' eigenstates of linearly independent generators of an internal SU(3) symmetry
and the `flavor' eigenstates of linearly independent generators of an internal SU(2) symmetry, $\mathbf{\Gamma}^0$ and $\mathbf{J}_3$
on $\bar{\psi}$ within the confines of the
extended Poincare group defining particle state representations necessarily involves CKM mixing between three `generations'.

The generators $\{ \tilde{\mathbf{t}}_1, ... , \tilde{\mathbf{t}}_8  \}$ define an independent group of SU(3) transformations
on the 4-spinors that will be denoted $\tilde{\mathbf{\mathcal{S}}}$.  Generally using this procedure for arbitrary CKM
transformations, an internal SU(3) symmetry group for the
transformed state spinor $\bar{\psi}$ is defined using the CKM transformation
\be
\mathbf{M}^{(3)} \equiv \mathbf{U}^{-1}_{CKM} \,  \tilde{\mathbf{\mathcal{S}}} \:  \mathbf{U}_{CKM} ,
\label{Eq:M3}
\ee
where the generators $\tilde{\mathbf{t}}_r$ of the symmetry group $\tilde{\mathbf{\mathcal{S}}}$ must form a linearly independent
set of hermitian matrices in SU(4) consistent with the extended Lorentz group transforming particle state representations
within that same group space.

\subsection{Transformation properties of causal fields}

Pairs of quantum fields that obey microscopic causality either commute or anti-commute for space-like separations
($\vec{y}-\vec{x}$) of the space-time coordinates of those fields (i.e. outside of the light cone).
The form of a causal spinor field\index{causal spinor field} that has
the expected properties under parity, time reversal, and charge conjugation is given by
\bea
\hat{\mathbf{\Psi}}_{(\gamma)}^{(\Gamma)}(\vec{x}) \equiv {1 \over \sqrt{2}}
\sum_{J, s_z } \int {m c^2 \, d^3 p \over \epsilon(\mathbf{p})} 
\left [  {e^{{i \over \hbar} ( \mathbf{p} \cdot \mathbf{x} - 
\epsilon (\mathbf{p}) \, t ) } \over
(2 \pi \hbar)^{3/2}} \, \mathbf{u}_{(\gamma)}^{(\Gamma)}(\vec{p},m,J,s_z)  \,
\hat{a}_{(\gamma)}^{(\Gamma)}(\vec{p},m,J,s_z)  + \right . \nonumber  \\ \left .
(-)^{J + s_z} \, {e^{-{i \over \hbar} ( \mathbf{p} \cdot \mathbf{x} - 
\epsilon (\mathbf{p}) \, t ) } \over
(2 \pi \hbar)^{3/2}} \, \mathbf{u}_{(-\gamma)}^{(\Gamma)}(\vec{p},m,J,-s_z)  \,
\hat{a}_{(-\gamma)}^{(\Gamma) \dagger}(\vec{p},m,J,s_z)
\right ] ,  \quad  \quad
\label{FinalCausalSpinorField}
\eea
in terms of the creation and annihilation operators for the particle states. 
The normalization has been chosen to have non-relativistic correspondence,
${m c^2 \, d^3 p \over \epsilon(\mathbf{p})} \rightarrow d^3 p$ for $p << m c$ .
 The fields are causal in that they anti-commute/commute outside of the light cone
according to whether the spin $J_{max}=\Gamma$ is a half-integer or an  integer, i.e.
\be
\left [ \hat{\mathbf{\Psi}}_{(\gamma)}^{(\Gamma)}(\vec{y}),
\hat{\mathbf{\Psi}}_{(\gamma)}^{(\Gamma)}(\vec{x})
\right ]_{\pm} =0 \textnormal{ for }
(\vec{y}-\vec{x}) \cdot (\vec{y}-\vec{x}) >0 \textnormal{, where }
\pm = -(-1)^{2 J} .
\ee 
Microscopic causality compels a well defined relationship of the contributions of spinor states
$\mathbf{u}_{(\gamma)}^{(\Gamma)}(\vec{p},m,J,s_z)$ to those of their causal partners
$\mathbf{u}_{(-\gamma)}^{(\Gamma)}(\vec{p},m,J,-s_z)$ in the construction of
a causal field in configuration space.
Under general Poincare transformations, the fields transform according to
\be
\hat{U} (\mathbf{\Lambda} , \vec{a})  \, 
\left [\hat{\mathbf{\Psi}}_{(\gamma)}^{(\Gamma)}(\vec{x}) \right ]_b \,
\hat{U}^\dagger (\mathbf{\Lambda} , \vec{a}) ~=~
\sum_{b'} \mathcal{D}_{b b'} ^{(\Gamma)} (\mathbf{\Lambda}^{-1}) \,
\left [\hat{\mathbf{\Psi}}_{(\gamma)}^{(\Gamma)} (\mathbf{\Lambda}\vec{x}+ \vec{a}) 
 \right ]_{b'} ,
\label{CausalSpinorPoincareTranformation}
\ee
where the matrices $\mathcal{D}_{b b'} ^{(\Gamma)} (\mathbf{\Lambda})$ form a finite-dimensional
representation of the Lorentz group of transformations $\Lambda$ \cite{WeinbergQTF}. 

In order to establish a relationship between a causal spinor field
$\mathbf{\Psi}$ and the transformed spinor $\bar{\psi}$,
a particular local Euclidean rotation $\mathbf{R}_{L}$ will be parameterized as follows:
\begin{displaymath}
\mathbf{R}_{14} = \left (
\begin{array}{cccc}
\cos(\zeta_{14}) & 0 & 0 & \sin(\zeta_{14}) e^{ i \omega_{14}} \\
0 & 1 & 0 & 0 \\
0 & 0 & 1 & 0  \\
-\sin(\zeta_{14}) e^{ -i \omega_{14}} & 0 & 0 & \cos(\zeta_{14}) 
\end{array}
\right ) , \quad
\mathbf{R}_{13} = \left (
\begin{array}{cccc}
\cos(\zeta_{13}) & 0 & \sin(\zeta_{13})  e^{ i \omega_{13}} & 0 \\
0 & 1 & 0 & 0 \\
- \sin(\zeta_{13})e^{ -i \omega_{13}} & 0 &  \cos(\zeta_{13}) & 0  \\
0 & 0 & 0 & 1 
\end{array}
\right ) , 
\end{displaymath}
\be
\mathbf{R}_{12} = \left (
\begin{array}{cccc}
\cos(\zeta_{12})  & \sin(\zeta_{12})  e^{ i \omega_{12}} & 0 & 0 \\
 -\sin(\zeta_{12})  e^{- i \omega_{12}} &  \cos(\zeta_{12})   & 0 &0 \\
0 & 0 & 1 & 0 \\
0 & 0 & 0 & 1
\end{array}
\right ) , \quad
\mathbf{R}_{L} \equiv \mathbf{R}_{14} \mathbf{R}_{13}\mathbf{R}_{12}
\ee
where the angles and phases are chosen according to the general form
\be
\mathbf{\Psi}(x) = \mathbf{R}_{L}(x) \bar{\psi}(x)= \left (
\begin{array}{l}
\phi_1 (x) e^{i \omega_1 (x)} \\
\phi_2 (x) e^{i \omega_2 (x)} \\
\phi_3 (x) e^{i \omega_3 (x)} \\
\phi_4 (x) e^{i \omega_4 (x)} 
\end{array}
\right ) \, = \,  \left (
\begin{array}{l}
\: \: \:  \phi_1 (x) e^{i \omega_1 (x)} \\
-\phi_1 (x) \tan(\zeta_{12}) \sec(\zeta_{13}) \sec(\zeta_{14})  e^{i \omega_2 (x)}\\
-\phi_1 (x) \tan(\zeta_{13})  \sec(\zeta_{14})  e^{i \omega_3 (x)}\\
-\phi_1 (x)  \tan(\zeta_{14})  e^{i \omega_4 (x)}
\end{array}
\right ),
\ee
with $\omega_{1s} \equiv \omega_1-\omega_s$.  The transformed spinor  with the internal
symmetry groups then has the form
\be
\bar{\psi}(x) = \left (
\begin{array}{c}
 \sqrt{(\phi_1(x))^2 + (\phi_2 (x))^2 + (\phi_3 (x))^2 + (\phi_4 (x))^2 }\:  e^{i \omega_1 (x)}\\
0 \\
0 \\
0
\end{array}
\right ) = \mathbf{R}_{L}^{-1}(x) \mathbf{\Psi}(x) .
\label{Eq:psibarfromPsi}
\ee
Using this relationship, internal local symmetries on the general causal fields will be
demonstrated in the next section.

\subsection{Local gauge symmetries of linear spinor fields}

Using the internal SU(3) symmetry transformation $\mathbf{M}^{(3)}$ on $\bar{\psi}$
from Eqn. \ref{Eq:M3}, and the relationship of the transformed spinor $\bar{\psi}$ to the general causal spinor
field expressed in Eqn. \ref{Eq:psibarfromPsi}, a  local internal SU(3) symmetry
on the causal spinor field is given by
\be
\mathbf{U}^{(3)} (x) =\mathbf{R}_L (x) \: \mathbf{M}^{(3)}(x) \:  \mathbf{R}_L^{-1}(x) = 
\mathbf{R}_L (x)  \mathbf{U}^{-1}_{CKM} \,  \tilde{\mathbf{\mathcal{S}}}(\underline{\alpha}(x)) \:  \mathbf{U}_{CKM}  \mathbf{R}_L^{-1}(x)  ,
\label{Eq:causalSU3}
\ee
where $\alpha^s(x)$ are the eight generally locally dependent gauge group parameters of SU(3), and one representation
of $\tilde{\mathbf{\mathcal{S}}}$ is given by $\tilde{\mathbf{\mathcal{S}}}(\underline{\alpha}) = e^{i \sum \alpha^s \tilde{\mathbf{t}}_s}$. 
It should be reemphasized that the set of SU(3) generators $ \tilde{\mathbf{t}}_s$ on the SU(4) space must be a subset
of the linearly independent generators that include those of the extended Lorentz group.

Similarly, the internal SU(2) symmetry transformation $\mathbf{M}^{(2)}$ on $\bar{\psi}$ from
Eqn. \ref{Eq:M2} defines a local internal SU(2) symmetry on the causal spinor field given by
\be
\mathbf{U}^{(2)} (x) =\mathbf{R}_L (x) \: \mathbf{M}^{(2)}(\underline{\theta}(x) ) \: \mathbf{R}_L^{-1}(x)  ,
\label{Eq:causalSU2}
\ee
where $\theta^j (x)$ are the three generally  locally dependent gauge group parameters of SU(2),
and  $\mathbf{M}^{(2)}(\underline{\theta}) = e^{i \sum \theta^j \tau_j}$. 
Both (\ref{Eq:causalSU3}) and (\ref{Eq:causalSU2}) are local internal symmetries on the causal spinor field
$\mathbf{U}(x) \mathbf{\Psi}(x) =\mathbf{\Psi}(x) $. 
 However,
the internal SU(2) and SU(3) symmetries do not share the same eigenbasis, since no set of linearly independent
generators including the extended Lorentz group, SU(2), and SU(3) can be found.  The different eigenbases are related via
CKM mixing in the enlarged unified group via $\mathbf{U}_{CKM}$.

For any physical system with an internal symmetry group, the assignment of local space-time coordinate
dependence to group transformation parameters results in a system with gauge invariance, as long
as the generators for space-time translations $\hat{P}_\beta$ are replaced in any Lagrangian
or field equation describing the dynamics using minimal coupling
$\hat{P}_\beta \rightarrow \mathbf{1} \hat{P}_\beta - \sum_r {q \over c} A_\beta^r (x) \mathbf{G}_r$,
where $\mathbf{G}_r$ represents the generator of infinitesimal transformations along group parameter $\alpha^r (x)$,
and $A_\beta^r (x)$ represents the gauge field.
The local internal symmetry $\mathbf{U}(\underline{\alpha}(x))$ is maintained as long as the gauge fields transform according to
(see reference \cite{JLHM85} or section 3.4.1 of reference \cite{JLFQG})
\be
A_\beta^r (x :\underline{\alpha}) = \sum_s A_\beta^s (x) \, \oplus_s^r (\underline{\alpha(x)}) + a_\beta^r(x) , \quad
\partial_\beta \mathbf{U}(\underline{\alpha}(x)) = {q \over \hbar c} \sum_s a_\beta^s (x) i \mathbf{G}_s \mathbf{U}(\underline{\alpha}(x)),
\ee
where $\mathbf{U}(\underline{\alpha}) \mathbf{G}_s \mathbf{U}^{-1}(\underline{\alpha}) \equiv \sum_r \oplus_s^r (\underline{\alpha}) \mathbf{G}_r$,
and the functions $a_\beta^s (x)$ are related to the derivative of the group parameters.
The gauge group topology of the local mapping in space-time of the group structure functions determines the monopole structure of
the sources of the gauge interactions (see section 4.2 of \cite{JLFQG}). 
The inclusion of geometrodynamics occurs via the principle of equivalence, 
with a replacement of the operations in locally flat space-time using covariance:
$\mathbf{\Gamma}^\mu P_{\xi^\mu} \rightarrow \mathbf{\Gamma}^\mu
{\partial x^\beta \over \partial \xi^\mu} P_\beta$, where $\xi^\mu$ are locally flat coordinates with conjugate momenta $P_{\xi^\mu}$.
The operators $P_{\xi^\mu}$ are the generators in the extended Poincare algebra.

\section{Conclusions}
\indent

General formulations of scattering theory that are unitary, maintain quantum linearity in space-time translation generators,
have positive definite energies, and have
straightforward cluster decomposition properties, can be constructed in
a straightforward manner using linear spinor fields. 
The fundamental representation of linear spinor fields unifies a set of internal local symmetries
including a U(1) symmetry along with 11 additional hermitian generators that can represent
a linearly independent SU(2) symmetry or a linearly independent
SU(3) symmetry, but not both.  The eigenbasis of the linearly independent SU(3) symmetry has been shown to
be related to that of the SU(2) symmetry via a CKM transformation mixing the symmetries in SU(4).
Internal local gauge symmetries for SU(2) and SU(3) on causal spinor fields have been demonstrated. 

The geometrodynamics of general relativity is directly incorporated in the
field equation satisfied by the linear spinor fields through the principle of equivalence. 
Furthermore, group algebraic invariants
explicitly include the Minkowski metric as calculated using the non-abelian
algebra of generators that carry space-time indeces, extending interior group structure to the dynamics
of general coordinate transformations. The piece of the group algebra 
that connects the group structure to metric gravitation via the equivalence principle (the algebra
of $\hat{\Gamma}^\beta$)
necessitates the inclusion of an additional group operator that generates affine parameter
translations for massless particles.  This allows dynamic mixing of massless particles
in a manner not allowed by the standard formulations of Dirac or Majorana.
On-going work is examining the extent that the present neutrino mixing phenomenology can be
modeled considering neutrinos as \emph{massless} spinor fields of differing transverse mass. 
Future work will examine the quantum number flows via the self-adjoint $(\gamma)=0$ degenerate
eigenstates of  $\hat{\Gamma}^\mu \: \hat{P}_\mu$ for the $\Gamma=1$ representation
of linear spinor fields that can mix with massless vector particles. 
This representation  contains a self-adjoint scalar particle, a self-adjoint vector particle,
another vector particle, and its adjoint vector particle.

\section{Acknowledgements}

The author wishes to acknowledge the support of Elnora Herod and
Penelope Brown during the intermediate periods prior to and after
his Peace Corps service (1984-1988), during which time the most significant aspects of this long-termed project were
accomplished.  This work was partially motivated by collaborations with H. Pierre Noyes at SLAC in developing
relativistic multiparticle unitary scattering formulations with cluster decomposability, and explorations of the
geometry of quantum flows with Harry Morrison at MIT and UC Berkeley. 
In addition, the author acknowledges the
hospitality of the Department of Physics at the University of Dar
Es Salaam during the nearly three years from 1985-1987 in which a substantial portion of
this work was done. The author also wishes to express his appreciation of the
hospitality of Lenny Susskind and the Stanford Institute for Theoretical Physics during his year of
sabbatical leave (2013), when some of the work on the internal gauge group structure 
was examined.  Finally, the author acknowledges the support of Mickey Chu and the Visiting
Faculty Program at Brookhaven National Laboratory, where initial comparisons of the mixing of
massless linear spinor fields and massive neutrino fields with known neutrino phenomenology were made.
This paper is a modified re-submission of arXiv:submit/1498541 first submitted 4 Mar 2016.


\begin{thebibliography}{999} 
\baselineskip=17pt 
\itemsep = 2pt 


\bibitem{JLFQG}
 J. Lindesay, \textit{Foundations of Quantum Gravity},
 Cambridge University Press  (2013), ISBN 978-1-107-00840-3.

\bibitem{LMNP} J.V. Lindesay, A.J. Markevich, H.P. Noyes, and G. Pastrana, 
Phys. Rev. D33, 2339 (1986).

\bibitem{AKLN} M. Alfred, P. Kwizera, J.V. Lindesay, and H.P. Noyes, 
A Non-Perturbative, Finite Particle Number Approach to Relativistic Scattering Theory, 
SLAC-PUB-8821, hep-th/0105241 (2001).

\bibitem{Dirac} P.A.M. Dirac,
Proc. Roy. Soc. (London), A117, 610 (1928);
ibid, A118, 351 (1928).

\bibitem{BjDrell} J.D. Bjorken and S.D. Drell,
\textit{Relativistic Quantum Mechanics}, McGraw-Hill, New York (1964).

\bibitem{JLLSF13}
J. Lindesay, ``Linear Spinor Fields in Relativistic Dynamics", arXiv:1312.0541 [hep-th] (2013) 13 pages.

\bibitem{Cabibbo}
N. Cabibbo, ``Unitary Symmetry and Leptonic Decays". \textit{Physical Review Letters} \textbf{10} (12): 531-533
(1963). doi:10.1103/PhysRevLett.10.531 .

\bibitem{KMmix}
M. Kobayashi and T. Maskawa, ``CP-Violation in the Renormalizable Theory of Weak Interaction".
\textit{Progress of Theoretical Physics} \textbf{49} (2): 652-657 (1973). doi:10.1143/PTP.49.652 .

\bibitem{WeinbergQTF}
S. Weinberg, \textit{The Quantum Theory of Fields}, Cambridge University Press,
Cambridge (1995).

\bibitem{JLHM85}
H. Morrison and J. Lindesay, ``The Geometry of Quantum Flow".
\textit{Mathematical Analysis of Physical Systems}, pp 135-167,   R. 
Mickens, editor.   Van Nostrand Reinhold, Co., New York  (1985)


\end{thebibliography}
\end{document}